\documentclass[prl,twocolumn,showpacs]{revtex4} \usepackage{epsfig} \usepackage[nooneline]{subfigure}
\begin{document}

\author{Daniela Froemberg and Igor M. Sokolov} \title{Stationary
spatial structures in reaction-subdiffusion} \date{\today}
\affiliation{Institut f\"ur Physik, Humboldt-Universit\"at zu Berlin,
Newtonstr. 15, D-12489, Berlin, Germany} \date{\today}


\begin{abstract} We discuss stationary concentrations of reactants in
an $\mathrm{A}+\mathrm{B} \rightarrow 0$ reaction under subdiffusion
and show that they are described by stationary reaction-diffusion
equations with a nonlinear diffusion term. We consider stationary
profiles of reactants' concentrations and of reaction zones in a flat
subdiffusive medium fed by reactants of both types on its both sides
(a subdiffusive gel reactor). The behavior of the concentrations and
of the reaction intensity in subdiffusion differs strikingly from
those in simple diffusion.  The most important differences correspond
to the existence of accumulation and depletion zones close to the
boundaries and to non-monotonous behavior of the reaction intensity
with respect to the strength of the minor source. The implications of
these results for other situations are also discussed.
\end{abstract}

\maketitle

Many phenomena in systems out of equilibrium are described as reactions between 
diffusing species. Apart from chemistry, examples are 
trapping and annihilation of excitons or recombination of charge carriers in physics, or
predator-prey relations in ecology. For normal diffusion
the situation is described in terms of reaction-diffusion
equations. Many systems however exhibit anomalous diffusion with a
mean square displacement scaling as
$\left<r^2(t)\right>\propto t^{\alpha}$, with $0<\alpha<1$ in the
subdiffusive case and $\alpha>1$ in the superdiffusive case, which is not described by 
a diffusion equation \cite{Phys2day,PhysWorld}.  
Subdiffusion can often be modeled within the
framework of continuous time random walks (CTRW) with a heavy-tailed
waiting time density function $\psi(t) \propto t^{-1-\alpha}$, 
which yields a fractional subdiffusion equation.  In
analogy to reaction-diffusion, fractional reaction-subdiffusion
equations have been proposed, where an additional fractional time
derivative acts either on the spacial Laplacian \cite{Wearnes,Koszt} or on both 
the spatial Laplacian
and the reaction term \cite{YAK,SWT}. Refs.\cite{SSS1,SSS2} showed that
the reaction-subdiffusion equations do not follow by simply changing a
diffusion operator in a reaction-diffusion equation for a subdiffusion
one. The situation here was pertinent to initial-condition problems.
A rather general approach to reaction-subdiffusion equations was
proposed in Ref.\cite{horsthemke}, where, after deriving the set of
reaction-subdiffusion equations the authors analyzed the properties of
Turing instability in corresponding reaction-subdiffusion systems. In
what follows we discuss the simplest boundary condition problem in
which nontrivial stationary spacial structures appear, namely the
irreversible $\mathrm{A} + \mathrm{B} \rightarrow 0$ reaction
\cite{YAK}, and focus on stationary concentration profiles under given,
steady concentrations of reactants on the boundaries (Ref.\cite{YAK}
started with one reactant, A or B, on each side). We show that the
spacial structures under reaction-subdiffusion are strikingly
different from ones emerging in reaction-diffusion.
We first give the derivation of subdiffusion-reaction equations and
stationary subdiffusion-reaction equations by generalizing the scheme
put forward in Refs.\cite{SSS1,SSS2}.  As in \cite{horsthemke},
subdiffusion is considered within a CTRW scheme, and the reaction
locally follows the mass action law. We then proceed by discussion of
stationary forms of concentrations' distributions and of those of
reaction zones. 

In a CTRW a particle arriving at a site $i$ at time
$t^{\prime}$ stays there for a sojourn time $t$, which is given by the
probability density function $\psi(t)$. Leaving the site it makes a step
with probability 1/2 in either direction. In the following discussion
we confine ourselves to the one-dimensional situation, however the
generalization to whatever other geometry is quite evident.

The generalized reaction-diffusion equations are based on two balance
conditions. The balance equation for A-particles at each site reads:
\begin{eqnarray} &&\dot{A}_i(t) = j_{i}^{+}(t)-j_{i}^{-}(t) +
R_i\{A,B\} \label{balance} \\ &&= \frac{1}{2} j_{i-1}^{-}(t)+
\frac{1}{2} j_{i+1}^{-}(t)- j_{i}^{-}(t)+ R_i\{A,B\}.
\label{balance0}
\end{eqnarray} where $j_{i}^{-}(t)$ is the loss flux of A-particles
at site $i$, i.e. the probability for an A-particle to leave $i$ per
unit time, $j_{i}^{+}(t)$ is the gain flux at the site, and
$R_i\{A,B\}=-kA_iB_i$ is a reaction term, describing particles' loss
due to reaction.  Since in our case the equations for A and B
particles are symmetric, we concentrate on the equations for the A
particles. A generalized reaction-diffusion equation is a combination
of the continuity equation, Eq.(\ref{balance0}), and the equation for
the loss fluxes $j^{-}(t)$ following from the assumption about the
distribution of sojourn times $\psi(t)$ and survival probability
$P(t,t_0)$.

According to the sojourn time distribution, the loss current for site
$i$ at time $t$ is connected to the gain current for the site at all
previous times and with the survival probability. Namely, the
particles which leave site $i$ at time $t$ (making a step from
$i$ to one of its neighbors) were either at site $i$ from the very
beginning (and survived), or arrived there at some time $t^{\prime
}<t$ and survived until $t$. The probability density to make a step at
time $t$, having arrived at $t^{\prime }$, is given by the waiting
time distribution $\psi(t-t^{\prime })$. We have then $j^{-}(t) =
\psi(t)P_i(t,0)A_i(0)+\int_{0}^{t}\psi(t-t^{\prime })
P_i(t,t')j_i^{+}(t^{\prime })dt^{\prime }$, which, by using
Eq.(\ref{balance}), can be rewritten in the form
\begin{eqnarray} j_{i}^{-}(t) = \psi(t)
P_i(t,0)A_{i}(0)+\int_{0}^{t}\psi (t-t^{\prime }) P_i(t,t^{\prime })
\times \nonumber \\ \times \left[ \dot{A}_{i}(t^{\prime
})+j_{i}^{-}(t^{\prime })+kA_i(t')B_i(t')\right] dt^{\prime }.
\label{balance2}
\end{eqnarray} The survival probability of A at $i$ is given by
the classical kinetic equation $\frac{d}{dt}P_i(t) = -kB_i (t)P_i(t)$
and depends on the time-dependent B-concentration via
\begin{equation} P_i(t,t_0) = \exp\left(-k\int_{t_0}^{t} B_i(t')dt'
\right).
\label{survi}
\end{equation}  
At this stage we also can assume the concentrations to be
slowly changing is space, and change to a continuous coordinate
$x=ai$, with $a$ being the lattice spacing:
\begin{equation} \dot{A}(x,t) = \frac{a^2}{2} \Delta j^{-}(x,t)
-kA(x,t)B(x,t). \label{fin1}
\end{equation} Eq.(\ref{fin1}) together with Eqs.(\ref{balance2}) and
(\ref{survi}) and their counterparts for B-concentration give us the
full system of equations for time-dependent concentrations. This
system can be transformed to a special case of reaction-subdiffusion
equations considered in Ref. \cite{horsthemke}. However, at this stage
it is easier to proceed with the equations in our form.

In our previous discussion we considered an initial condition problem,
where A-particles were introduced at time $t=0$ into the system and
followed the evolution of their concentration. In the case of normal
diffusion the situation in steady state (achieved, say, if the
concentration of the particles at boundaries of the system is fixed by
external sources) is given by the same
reaction-diffusion equations, with the time derivatives at the left
hand side put to zero. For the case of subdiffusion the situation is a bit
more involved. 

Let us assume that in the course of time the system achieves a steady
state characterized by constant concentrations $A(x)$ and $B(x)$. Such
a steady state is maintained through the particles' sources at the
boundaries of the system, no particle sources exist in the
interior. Let us label the particles according to the time $t_0$ they
were introduced into the system, so that e.g. $A(x,t|t_0)dt_0$ is the
concentration at point $x$ at time $t$ of A-particles introduced between $t_0$ and $t_0 + dt_0$
(a partial concentration of A). The partial
concentration $A(x,t_0|t_0)$ of newly introduced particles is zero
everywhere in the interior of the system. The overall concentration of
A-particles at site $x$ is given by the integral
\begin{equation} A(x)= \int_{-\infty}^t A(x,t|t_0) dt_0.
\label{DefA}
\end{equation} In a steady state $A(x,t|t_0)$ can only be a function
of the difference of the time-arguments, i.e. of the elapsed time
$t_e=t-t_0$ so that $A(x,t|t_0) = A(x,t-t_0)$ and the equations for
the partial concentrations $A(x,t|t_0)$ are given by Eqs.(\ref{fin1})
with $t$ changed to $t_e$. Moreover, the overall concentration $A(x)$
is given by $A(x)= \int_0^\infty A(x,t_e) dt_e$.  Since $A(x)$ and
$B(x)$ are time-independent, the survival probabilities $P_A(x,t,t')$
and $P_B(x,t,t')$ in reaction-subdiffusion equations are the functions
of the differences of their time-arguments so that
$P_A(x,t,t')=\exp\left[-k B(x)(t-t')\right]$.  The integral in the
equation for the flux now takes the form of a convolution:
\begin{eqnarray}
&&j^{-}(x,t|t_0) = \psi(t) P_A(x,t-t_0) A(x,t_0|t_0) \nonumber \\
&& + \int_{t_0}^{t}\psi(t-t^\prime)P_A(x,t-t^{\prime }) \times \\
&& \times \left[ \dot{A}(x,t^{\prime}|t_0)+j^{-}(x,t^{\prime }|t_0) 
+ kA(x,t'|t_0)B(x) \right] dt^{\prime} \nonumber
\end{eqnarray}
where $j^{-}(x,t^{\prime }|t_0)$ are the loss fluxes for those
A-particles which were introduced into the system at time $t_0$. We
now pass to the Laplace domain with respect to $t_e$ and denote
$\tilde{A}(x,u) = \int_{t_0}^\infty A(x,t|t_0) \exp[-u(t-t_0)]dt$. The
Laplace transform of the product $\Psi(t,x)=\psi (t) \exp\left[-k
B(x)t\right]$ is given by the shift theorem and is equal to
$\tilde{\Psi}(u,x)=\tilde{\psi}(u+kB(x))$, so that
\begin{equation} \tilde{j}^-(x,u) =
\frac{[u+kB(x)]\tilde{\psi}(u+kB(x))}{1-\tilde{\psi}(u+kB(x))}
\tilde{A}(x,u).
\label{LaID}
\end{equation} Inserting this into the first equation we get in the
Laplace domain
\begin{eqnarray} && u \tilde{A}(x,u) - A(x,t_0|t_0) = \\ &&
\frac{a^2}{2} \Delta \frac{[u+kB(x)]
\tilde{\psi}(u+kB(x))}{1-\tilde{\psi}(u+kB(x))} \tilde{A}(x,u) -k
\tilde{A}(x,u) B(x), \nonumber
\label{EqforA}
\end{eqnarray} where $A(x,t_0|t_0)$ differs from zero only at the
boundaries. With $A(x)=\tilde{A}(x,0)$ the stationary
concentration $A(x)$ in the interior of the system is given by
\begin{equation} \frac{a^2}{2} \Delta \frac{kB(x)
\tilde{\psi}(kB(x))}{1-\tilde{\psi}(kB(x))} A(x) -k A(x) B(x) = 0,
\label{concA}
\end{equation} with the boundary conditions corresponding to the given
concentrations on the boundaries. For a Markovian case of regular diffusion,
corresponding to $\psi(t)=\tau^{-1} \exp(-t/\tau)$, one has $\tilde{\psi}(u) = 1/(1+u\tau)$ 
so that this equation reduces to $(a^2/2 \tau)
\Delta A(x) -k A(x) B(x) = 0$, a usual stationary reaction-diffusion equation. In
the non-Markovian case, corresponding to subdiffusion, the waiting time distribution in the Laplace
domain is $\tilde{\psi}(u) \simeq 1-(\tau u)^{\alpha} \Gamma
(1-\alpha)$ for small $u$ and Eq.(\ref{concA}) reads
\begin{equation} \frac{a^2}{2} \frac{1}{\tau^{\alpha} \Gamma
(1-\alpha)} \Delta B(x)^{1-\alpha} A(x) -k^\alpha A(x) B(x) =
0. \label{nMarkAB}
\end{equation} 
It is easy to see that the Markovian equation is a
special case of Eq.(\ref{nMarkAB}) for $\alpha =1$.  The combination
$D_\alpha = a^2/2 \tau^\alpha \Gamma(1-\alpha)$ stands for a
(generalized) diffusion coefficient. The full system of steady state
equations is given by Eq.(\ref{concA}) and the corresponding equation
for B. It is interesting to stress that the system of equations with
additional temporal operator acting on the Laplacian in the case of
initial-condition problem turns to a system of \textit{nonlinear}
reaction-diffusion equations with nonlinearity in the diffusion term
for a stationary boundary-condition problem.

\begin{figure}[b]
\centering
{\includegraphics[width=0.5\textwidth]{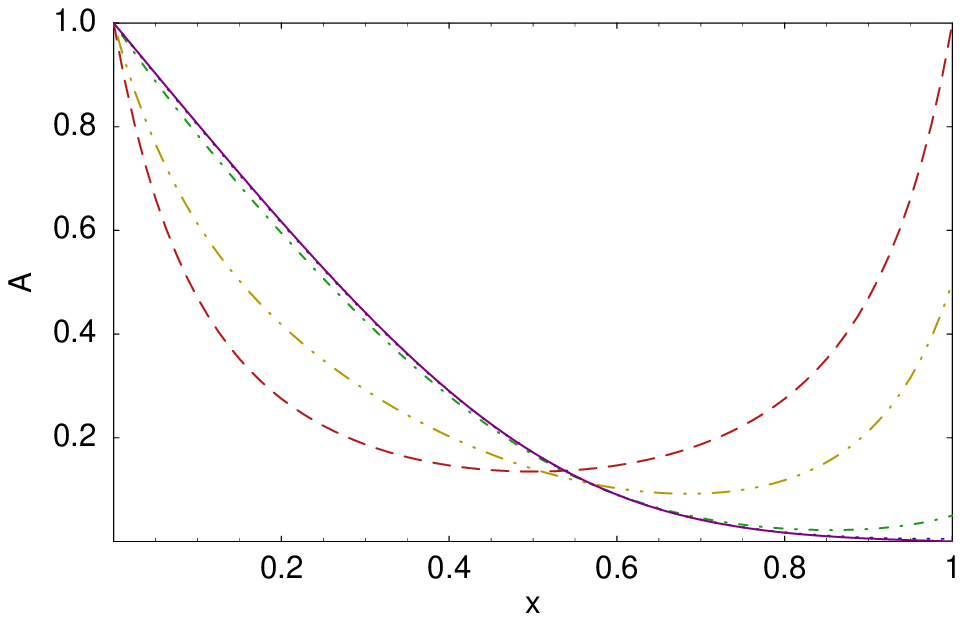}} \\
\bigskip
{\includegraphics[width=0.5\textwidth]{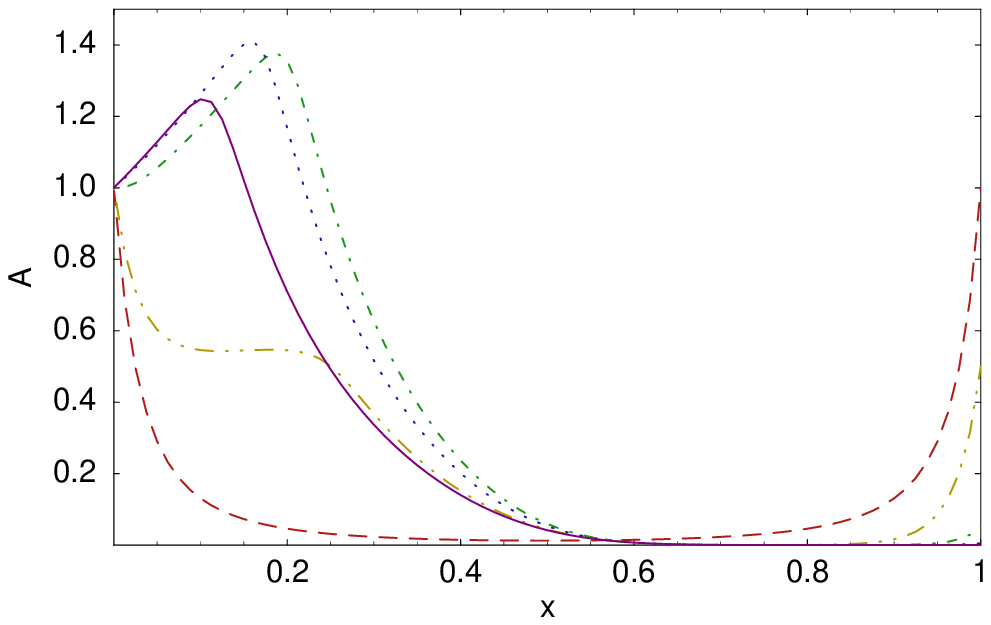}}
\caption{\label{bildA} Stationary particle concentration profile
$A(x)$ for $\alpha=1$ (upper panel) and for $\alpha=0.9$ (lower panel);
$A(1)=1$ (dashed), $0.5$ (dash-dot-dot), $5\cdot10^{-2}$ (dash-dot),
$5\cdot10^{-3}$ (dotted), and $1\cdot10^{-4}$ (solid line), see text for details. Note
the growth of the peak and its subsequent decay and outwards motion
with decreasing minor source strength for the subdiffusive case; the behavior of
the particle concentration profile in regular diffusion is monotonous with respect
to the strength of the minor source and tends to a
limiting form for $A(1) \rightarrow 0$.}
\end{figure}

As an example we consider a system on an interval $(0,1)$ with given
reactants' concentrations on the boundaries of the interval. The
physical system we have in mind is a gel reactor or a porous medium in
contact with two well-mixed reservoirs on both sides. The
concentrations $A(0)=B(1)=1$ at the major sources are fixed. For the
sake of simplicity we consider here symmetric situation with $A(0)=B(1)$ and
$A(1)=B(0)$. Due to
symmetry the $B$-concentration is then always symmetric to the
$A$-concentration, i.e. $B(x)=A(1-x)$. The concentration $B(0)=A(1)$ 
will be called the minor source strengths (except for the symmetric case
$A(1)=B(0)=1$). In Fig.1 we show numerical
results for the steady-state Eqs.(\ref{nMarkAB}) obtained by
semi-implicit relaxation algorithm. The results for subdiffusion (here
with moderate $\alpha=0.9$) are compared to the ones for normal
diffusion ($\alpha = 1$ and hence linear diffusion operator). Shown is
the behavior of the A-concentrations and the one of the reaction
intensity $kA(x)B(x)$. The parameters are: $k = 0.01$, $D_\alpha =
1/2\Gamma(1-\alpha)$ for $\alpha=0.9$ and $D_\alpha = 1/2$ for
$\alpha=1$. The $A(0)=B(1)$ concentration is fixed to be $A(0)=1$, the
other concentration varies from $B(0)=A(1)=1$ (symmetric case, when
the CTRW-reactor separates two stoichiometric reacting mixtures) to
$B(0)=A(1)=10^{-4}$.

\begin{figure}[b]
\centering
{\includegraphics[width=0.5\textwidth]{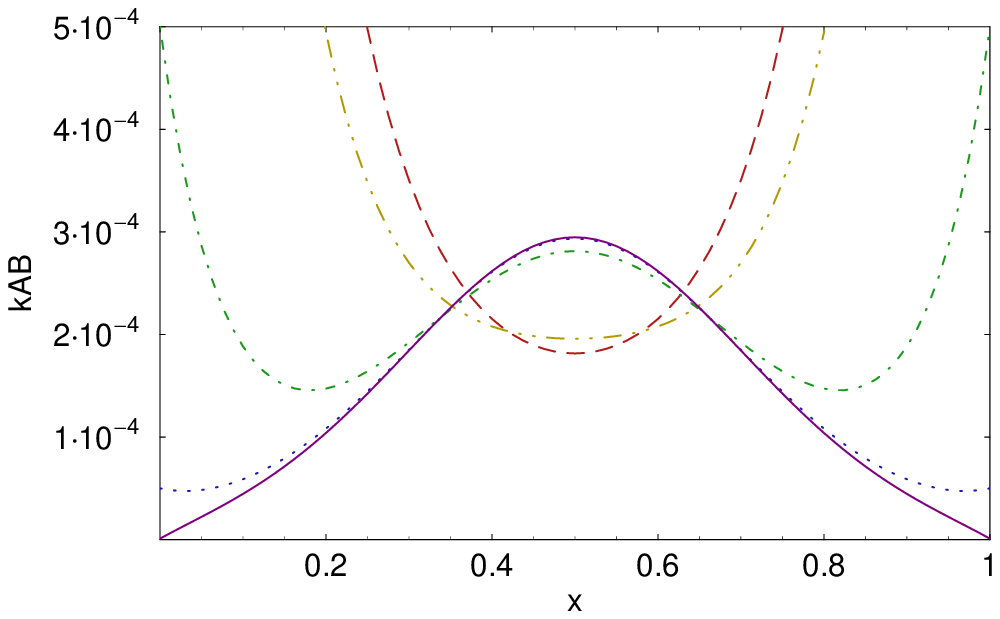}} \\
\bigskip
{\includegraphics[width=0.5\textwidth]{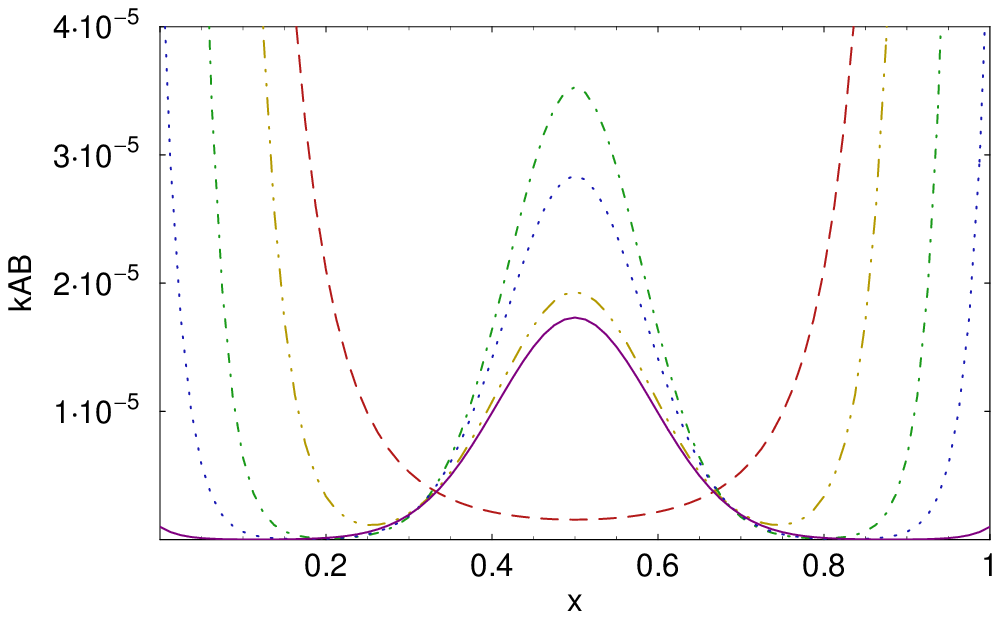}}
\caption{\label{bildR} Stationary reaction intensity profile $kA(x)B(x)$ for $\alpha=1$ (upper panel) and 
for $\alpha=0.9$ (lower panel). The parameters are the same as in Fig.1, namely: $A(1)=1$ (dashed line)
, $0.5$ (dash-dot-dot), $5\cdot{10^{-2}}$ (dash-dot), $5\cdot{10^{-3}}$ (dotted),and $1\cdot{10^{-4}}$ (solid line).
In the diffusive case the stationary reaction intensity profile approaches a limiting form 
under decreasing minor source strength. For subdiffusion-reaction the height of the profile shows nonmonotonous behavior with respect to this parameter. Note the difference in vertical scales!}
\end{figure}

In the symmetric case both reaction-diffusion and
reaction-subdiffusion situations correspond to a very similar behavior
with maximal concentrations in the regions close to the boundaries
where the system is fed by reactants. For asymmetric boundary
conditions however the behaviors of the concentrations 
in regular diffusion and in subdiffusion differ strongly. One of the most marked
differences corresponds to a strong non-monotonicity of the
concentration close to the major source (side with the higher
concentration of A) indicating for accumulation of A particles in the
interior of the subdiffusive medium. Its counterpart on the other side
of the system is a depletion zone (corresponding to the symmetric
accumulation zone for B). The peak and the depletion zone are much
more pronounced for smaller $\alpha$; the choice of moderate
$\alpha=0.9$ was caused by our whish to show the behavior pertinent
to diffusion and to subdiffusion on the same scale. It is important to
note that the dependence of the height of the
accumulation peak on the strength of the minor source is
nonmonotonous: The reduction of the minor source strength $A(1)=B(0)$ leads
first to its growth, and then to its outwards
motion accompanied by decay.

Let us now turn to reaction intensities, Fig. 2. For the symmetric
case the reaction takes place mostly close to the boundaries of the
system.  For smaller $A(1)$ the reaction zone starts to form in the
middle of the system. However, also here striking differences between
the reaction-diffusion and the reaction-subdiffusion cases are
seen. In the reaction-diffusion case the dependence of the form of the
reaction zone on $A(1)$ is weak for small $A(1)$, and there exists a
clear limiting form for $A(1)=0$. This behavior is known and is used
in the time-scale separation approach of Refs.\cite{Sokolov,GalfiRatz}
based on the quasistatic approximation. For reaction-subdiffusion
the behavior of the reaction zone with respect to its height is
non-monotonous. When lowering $A(1)$, the maximum of reaction
intensity first gets higher and then starts to lower; and the
distribution as a whole broadens. The reason for this is quite
evident. The stationary reaction zone exists only if it is fed by A-
and B-reactants on the corresponding sides. Both in the diffusion and
in the subdiffusion case the reaction zone is the higher and the
narrower the larger is the particles' inflow into the reaction area.
This inflow is governed by the effective diffusion coefficient of the
corresponding reactants, which, in the subdiffusive case, depends on
the concentration of the reacting counterpart.  Since the effective
mobility of subdiffusing species decays in the course of time (the
number of steps per unit time goes as $t^{\alpha-1}$), the effective
diffusion in subdiffusion is caused by the reaction itself: it
corresponds to a reaction-induced diffusion term, just as in the
reactions between immobile species \cite{PostnikovSokolov,
Sander}. For zero $A(1)=B(0)$ concentration the effective diffusion
coefficient vanishes on the corresponding side of the system
preventing the inflow of reactants from their major sources into the
interior of the system. The reaction zone blurs and fades out. In this
case no stationary front exists. This effect is also clearly seen when
considering the time evolution of concentrations which can be done by
discussing the properties of the inverse Laplace transform of
Eq.(\ref{EqforA}). This means that the adiabatic approximation of
Refs.\cite{Sokolov,GalfiRatz} fails in subdiffusion, and the analysis
of the front's motion in this case has to be done anew.

Let us summarize our findings. We discussed the stationary form of
reactants' concentrations and of reaction zones in the
$\mathrm{A}+\mathrm{B} \rightarrow 0$ reaction in a subdiffusive
medium fed by reactants on its both sides. 
We show that the behavior of the concentration and of the
reaction intensity profiles in subdiffusion differs
strikingly from those in simple diffusion. The most important
differences correspond to the existence of accumulation and depletion
zones close to the boundaries and to non-monotonous behavior of the
reaction intensity with respect to the strength of the minor
source. These results have implications for the time-dependent front
forms in $\mathrm{A}+\mathrm{B} \rightarrow 0$ reaction, since the
adiabatic approximation used in the reaction-diffusion case would fail in
the reaction-subdiffusion one.

\end{document}